\documentclass[checkout,showpacs,aps,jap,twocolumn]{revtex4}

\newcommand{\bn}{\begin{enumerate}}
\newcommand{\en}{\end{enumerate}}
\newcommand{\ba}{\begin{eqnarray}}
\newcommand{\ea}{\end{eqnarray}}

\usepackage{graphicx,color}

\newcommand{\be}{\begin{equation}}
\newcommand{\ee}{\end{equation}}

\newcommand{\et}{{\it et al. }}

\newcommand{\clr}{\color{black}}

\def\prl{{ Phys. Rev. Lett. }}

\def\prb{{ Phys. Rev. B }}

\begin{document}

%\title{ Theory of ultrafast demagnetization:\\
%Perspectives from  spin-orbit-coupled Heisenberg system}

%\title{Demagnetization in a spin-orbit-coupled Heisenberg
%  system:\\ Applications to ultrafast spin dynamics}

%\title{Simple picture of demagnetization from a spin-orbit-coupled
%  Heisenberg system: From static to dynamic}

%\title{Essence of femtosecond demagnetization:\\ Exact results from a
%  spin-orbit-coupled Heisenberg system}

%\title{Understanding femtosecond magnetism:\\ A simple picture
%  from a spin-orbit-coupled Heisenberg system}

%\title{Exchange interaction  in femtomagnetism}
%\title{Electron exchange-interaction collapse as an alternative
%  mechanism for femtomagnetism }

%\title{A path to the consistent theory of femtosecond
%  magnetism:\\ Spin-orbit-coupled Heisenberg exchange model
%}

\title{An attempt to simulate laser-induced all-optical spin
  switching in a crystalline ferrimagnet}

\author{G. P. Zhang\footnote{Author to whom correspondence should be addressed.}} \affiliation{Department of Physics, Indiana
  State University, Terre Haute, IN 47809, USA}

\email{guo-ping.zhang@outlook.com.}

%\author{Nicholas Allbritton} \affiliation{Department
%  of Physics, Indiana State University, Terre Haute, IN 47809,
%  USA}

\author{Robert Meadows} \affiliation{Department
  of Physics, Indiana State University, Terre Haute, IN 47809,
  USA}

\author{Antonio Tamayo} \affiliation{Department
  of Physics, Indiana State University, Terre Haute, IN 47809,
  USA}

\author{Y. H. Bai} \affiliation{Office of Information
  Technology, Indiana State University, Terre Haute, Indiana 47809,
  USA}

 \author{Thomas F. George} \affiliation{Departments of Chemistry \&
   Biochemistry and Physics \& Astronomy \\University of
   Missouri-St. Louis, St.  Louis, MO 63121, USA }

\date{\today}

\begin{abstract}
{Interest in all-optical spin switching (AOS) is growing
  rapidly. The recent discovery of AOS in Mn$_2$RuGa provides a much
  needed clean case of crystalline ferrimagnets for theoretical
  simulations. Here, we attempt to simulate it using the
  state-of-the-art first-principles method combined with the
  Heisenberg exchange model. We first compute the spin moments at two
  inequivalent manganese sites and then feed them into our model
  Hamiltonian. We employ an ultrafast laser pulse to switch the
  spins. We find that there is a similar optimal laser field amplitude
  to switch spins.  However, we find that the exchange interaction has
  a significant effect on the system switchability. Weakening the
  exchange interaction could make the system unswitchable. This
  provides a crucial insight into the switching mechanism in
  ferrimagnets.  }
\end{abstract}


%ultrafast magnetization dynamics, 75.78.Jp

%75.40.Gb, 78.20.Ls, 75.70.-i, 78.47.J-}

%\keywords{femtomagnetism; exchange interaction}
 \maketitle

\newcommand{\tm}{\tau_m}
\newcommand{\ub}{\mu_{\rm B}}

Central to the magnetic storage device is the writing/reading speed of
magnetic bits in a storage medium. Traditionally, these operations are
mostly driven by an external magnetic field.  A full-optical driven
spin manipulation could break the speed barrier of several hundred
picoseconds set by the Zeeman interaction and magnetic dipole-dipole
interaction. In 1996, Beaurepaire \et \cite{eric} showed that when
they shone a 60-fs pulse on the ferromagnetic nickel thin film, they
found a sharp decrease in the Kerr signal within 1 ps. This finding
received immediate attention worldwide, and a new research field,
femtomagnetism, was born \cite{ourreview,rasingreview}. Research
intensified, and is far beyond the scope of the original research
interest. In 2007, Stanciu \et \cite{stanciu2007} showed that a
left-circularly polarized laser pulse can switch an up-spin to down,
while a right-circularly polarized laser pulse can switch a down-spin
up. This remarkable property represents an interesting new magnetic
phenomenon on an ultrafast time scale, although their compound,
GdFeCo, is not new. GdFeCo has been used in traditional
magneto-optical recording (see the references cited in
\cite{mplb16,mplb18}). However, being to able to switch spins on a
picosecond time scale optically is new, and has raised the possibility
for a real application. However, it is unclear how the laser pulse can
switch spins directly. Ostler \et \cite{ostler2012} further showed
that if the laser intensity is increased above a certain level,
regardless of laser helicity, each pulse can flip spins from one
direction to another deterministically. They argued that there is a
threshold intensity that one has to exceed to change from all-optical
helicity dependent spin switching to all-optical helicity independent
switching, but the actual picture is more complicated
\cite{jpcm11,jpcm13}.

For a long time, GdFeCo was the only material that shows AOS. Soon,
many more materials were found
\cite{mangin2014,hassdenteufel2014,schubert2014a,alebrand2014,lambert2014}. However,
these materials are mostly amorphous, which introduces an uncertainty
in theoretical simulations and represents a formidable task.  In 2017,
Vomir \et \cite{vomir2017} reported the first observation of AOS in a
Pt/Co/Pt ferromagnetic stack, but the switching is not complete. Very
recently, Banerjee \et \cite{banerjee2019} showed single-pulse
all-optical toggle switching of magnetization in
Mn$_2$RuGa. Mn$_2$RuGa is a ferrimagnetic Heusler compound, with a
cubic structure. Two Mn atoms are not equivalent, and have different
spin moments.  They are antiferromagnetically coupled. As shown before
\cite{prb17}, ferrimagnets have a big advantage over ferromagnets and
antiferromagnets.  This offers an ideal theoretical model.

\newcommand{\br}{{\bf r}}

\newcommand{\ik}{i{\bf k}}

\newcommand{\jk}{j{\bf k}}

\newcommand{\lk}{l{\bf k}}

\newcommand{\bk}{{\bf k}}

%\section{First-principles calculation}

\newcommand{\iik}{i,i,{\bf k}}

In this paper, we investigate all-optical switching in
Mn$_2$RuGa. Different from prior studies, we compute the spin moments
at two Mn sites using the first-principles density functional
theory. These spin moments are fed into the Heisenberg exchange model
with both spin-orbit coupling and a harmonic potential
\cite{epl15,epl16}.
We find that not any
arbitrary laser field amplitude can switch spins. There is a narrow
window of opportunity where the spins at two Mn sites can be switched
into their respective opposite directions. Because of the strong spin
moments at two Mn sites, its switching is very stable. Quite different
from other systems, we find that if we reduce the exchange
interaction, the spins precess strongly at both Mn sites, very much
like a regular antiferromagnet, instead of a ferrimagnet. These strong
spin oscillation occur at both Mn$_1$ and Mn$_2$ sites. Their
oscillation period is inversely proportional to the laser field
amplitude, similar to the Rabi frequency in a two-level system. For a
weak exchange interaction case, regardless of the magnitude of the
laser field amplitude, the spin switching is not observed. This points
out an entirely different scenario from ferromagnetic cases
\cite{epl16} and demagnetization \cite{jap19}. The rich picture that
is found here reveals a crucial effect of the effect of exchange
interaction on AOS, and should motivate further experimental and
theoretical investigations in the future.

Mn$_2$RuGa is a Heusler ferrimagnet, with a stoichiometric composition
of $X_2YZ$ and space group $F\bar{4}3m$. Two Mn atoms, $\rm Mn_1$ and
$\rm Mn_2$, are situated at $(4a)$ and $(4c)$, which are magnetically
inequivalent \cite{galanakis2014,yang2015}. Their spins are
antiferromagnetically coupled.  To properly investigate magnetic
properties of Mn$_2$RuGa, we employ the density functional theory
using the full-potential augmented plane wave method as implemented in
the Wien2k code \cite{wien2k,jpcm16}.
Our first-principles calculation shows that $\rm Mn_1$ has a spin
moment of 3.17232 $\mu_B$ and $\rm Mn_2$ has -2.30765 $\mu_B$. This
agrees with prior calculations \cite{galanakis2014,yang2015}.  So each
cell has a net spin moment of 1.02394 $\mu_B$, close to unity, which is
consistent with the nature of a stoichiometric half-metal
\cite{kurt2014}.  The spin moments on Ru and Ga are very small, and
will be ignored below.

It is not often recognized that the large spin moments on Mn atoms are
advantageous since the spin-orbit torque is proportional to the spin
moment \cite{epl15,epl16}.  Since it is not possible to simulate
all-optical spin reversal at the first-principles level, in the
following we will feed these two spin moments into our
Heisenberg-exchange coupled harmonic model
\cite{epl16,prb17,jpcm17a,jpcm17b}, and limit ourselves to a small
system with 101 lattice sites along the $x$ axis and $y$ axis,
respectively, with two monolayers along the $z$ axis.  Our Hamiltonian
is \begin{widetext}
\be H=\sum_i \left [\frac{{\bf p}_i^2}{2m}+V({\bf r}_i) +\lambda
  {\bf L}_i\cdot {\bf S}_i -e {\bf E}({\bf r}, t) \cdot {\bf
    r}_i\right ]-\sum_{ij}J_{ex}{\bf S}_i\cdot {\bf
  S}_{j}, \label{ham} \ee 
 \end{widetext}
where terms from the left to right are
respectively the kinetic energy operator of the electron, the
potential energy operator, the spin-orbit coupling, the interaction
between the laser and system, and the exchange interaction between
spins.  {\clr Our exchange parameter $J_{ex}$ is still time-independent,
although prior studies have shown that the exchange interaction itself
could be affected by the electric field
\cite{mentink2015,mikhaylovskiy2015}}.  $\lambda$ is the spin-orbit
coupling constant, $ {\bf L}_i$ and $ {\bf S}_i $ are the orbital and
spin angular momenta at site $i$, respectively, and {\bf p} and {\bf
  r} are the momentum and position operators of the electron,
respectively. We choose a spherical harmonic potential $V({\bf
  r}_i)=\frac{1}{2}m\Omega^2 {\bf r}_i^2$ with system frequency
$\Omega$.  ${\bf E}({\bf r}, t)$ is the laser field.  This model is
the only magnetic field-free model currently available to simulate
spin reversal, while the commonly used model employs an effective
magnetic field \cite{ostler2012}, which should be avoided. It
represents a small step towards a complete model.

\begin{figure}
\includegraphics[angle=0,width=0.8\columnwidth]{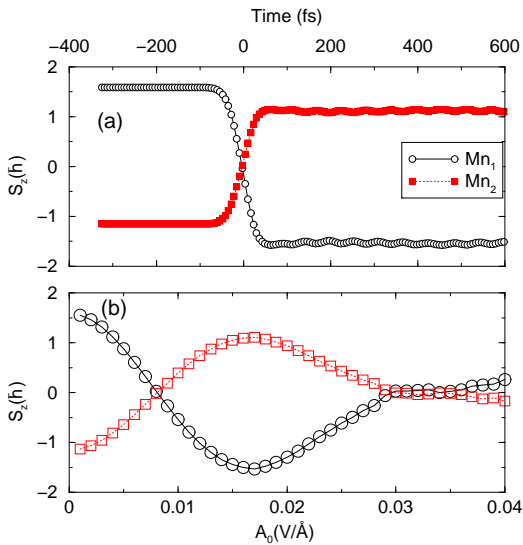}
\caption{(a) The $z$ component of the spins at the Mn$_1$ and Mn$_2$ sites
  as a function of time at the optimal laser field amplitude. Here, the
laser  amplitude is 0.017 $\rm V/\AA$, and the pulse duration is 60
fs. The empty circles denote the spin at the Mn$_1$ site, while the boxes refer
to the spin at the Mn$_2$ site. We see there is a clear spin reversal
upon laser excitation. (b) Dependence of the $S_z$ as a function of
the laser field amplitude $A_0$. The empty circles refer to the spin at
Mn$_1$ site, and the boxes refer to the spin at
Mn$_2$ site. A weak laser field does not reverse the spins, but a too
strong laser field can not either. There is a narrow window that one
can switch spins.
}
\label{fig1}
\end{figure}

In order to compute the spin change, we solve the Heisenberg equation
of motion \cite{jap19} for each spin operator at every site under
laser excitation \cite{prb17}. Figure \ref{fig1}(a) shows the spin
$z$ component at two Mn sites as a function of time. We employ a laser
pulse of 60 fs, with a field amplitude of 0.017 $\rm V/\AA$. We see
that the spin at the Mn$_1$ site starts from the positive $z$ axis (see
the circles). Upon laser excitation, it switches over the negative $z$
axis, while the spin at the Mn$_2$ site switches up from its $-z$
direction. This is consistent with the experimental observation
\cite{banerjee2019}. The strong spin moment stabilizes the entire
switching process. However, not any arbitrary laser field amplitude
can lead to faithful switching. Figure \ref{fig1}(b) shows how the
final spin changes with the laser field amplitude. The dependence is
highly nonlinear.  If we use a weak laser pulse, there is little
change in spins at both sites. But if the laser field is too strong,
the spins overturn toward the $xy$ plane, so there is no spin reversal
either. We find that the optimal field amplitude is 0.017 $\rm V/\AA$,
whose result is shown in Fig. \ref{fig1}(a). In this regard,
Mn$_2$RuGa is pretty much similar to other ferrimagnets where there is
an optimal amplitude \cite{epl16,prb17}. Microscopically, the real
situation is more complicated.

{\clr To this end, there is no generic understanding of spin switching
  in both ferromagnets \cite{vomir2017,lambert2014} and ferrimagnets
  \cite{stanciu2007}.  It has been often argued that the angular
  momentum exchange between two spin sublattices in ferrimagnetic
  GdFeCo \cite{mentink2012} is the key to AOS.  Theoretically, this
  momentum exchange picture is interesting, but such momentum exchange
  between sublattices, if it exists, occurs all the time through the
  exchange interaction, with or without the laser.  In other words, it
  must be something extra due to the laser that switches the spin. In
  GdFeCo, a rigorous testing is difficult because it is amorphous, and
  it is difficult to tell whether a model system really represents a
  true GdFeCo sample. This brings ambiguity to a theoretical
  simulation.  Mn$_2$RuGa removes this ambiguity completely.}

\begin{figure}
\includegraphics[angle=0,width=0.8\columnwidth]{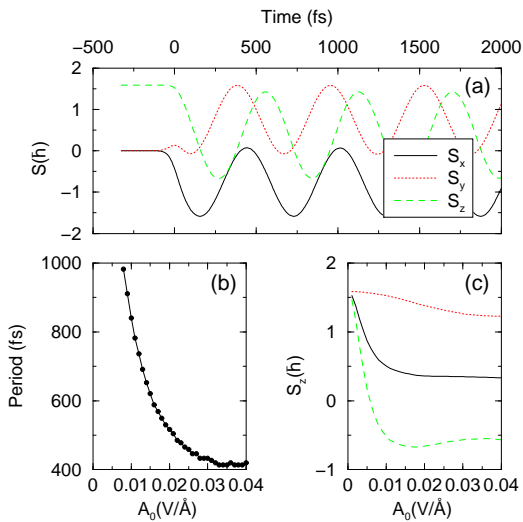}

\caption{ (a) Spin precession at Mn$_1$ site under a reduced exchange
  interaction. Here we choose $J=0.001$ eV. The rest of parameters are
  the same as those in Fig. \ref{fig1}. The solid, dotted, and dashed
  lines denote the $x$, $y$, and $z$ components of the spin,
  respectively. (b) The spin oscillation period decreases with laser
  field amplitude $A_0$.  (c) At any of the laser field amplitudes, spin
  reversal is not found. Only a strong oscillation is noticed. The
  solid line is the time-average of the spin, and the dotted and dashed
  lines refer to the maximum and minimum spin values, respectively. The
effect of the exchange interaction on spin reversal is much more
pronounced in Mn$_2$RuGa than in other materials. }
\label{fig2}
\end{figure}

As a first test, we investigate the effect of the exchange interaction on
AOS. We reduce the exchange interaction from 0.1 eV to 0.001 eV. From
prior studies, we know such a reduction does not constitute a major
issue for demagnetization in a system with a small spin moment
\cite{jap19}. Figure \ref{fig2}(a) shows the spin change at Mn$_1$
(the situation is similar at Mn$_2$), where the solid, dotted, and
dashed lines denote the $x$, $y$, and $z$ components, respectively.
Both the laser duration and amplitude are exactly the same as those in
Fig. \ref{fig1}. We see that there is a strong oscillation in all
these three components. We note in passing that these three components
must obey the operator permutation \cite{jpcm11}, where they can not
be considered a linear reversal \cite{vahaplar2009}. In principle, we
need to cut off the simulation around 1 ps, after which we need to
introduce damping, but to demonstrate the high accuracy of our
calculation, we do not use the damping.  These strong oscillations
resemble a pure antiferromagnetic case. The spins at two neighboring
sites are out of phase and remain antiferromagnetically coupled, even
upon laser excitation. The laser pulse essentially initiates the spin
dynamics, and the exchange interaction takes over, without switching
the spins.  For this reason, the angular momentum exchange picture for
AOS can not explain this even in the same ferrimagnet. The period of
the oscillation is not determined by the exchange interaction and spin
moment alone.  Figure \ref{fig2}(b) shows that as we increase the
laser field amplitude, the period becomes shorter. The small
fluctuation at the largest amplitudes is due to the period sampling
because the oscillation is not strictly harmonic. This laser-field
dependence of the oscillation period is very similar to the Rabi
period dependence. For all the field amplitudes that we investigate,
we do not see a case where the spins are reversed. Figure
\ref{fig2}(c) illustrates the average (solid line), maximum
(dotted line), and minimum (dashed line) of the final spin. We see the
average spin never becomes negative (the initial spin is along the
$+z$ axis). The maximum and minimum values show the limits of spin.
Our results point out an important fact: In a ferrimagnet, the effect
of the exchange interaction is far more complicated than thought.

Now, we have two cases: One shows AOS, and the other does not. We can
directly check whether the prior criteria proposed by Mentink \et
\cite{mentink2012} apply to them. Their argument is based on a
two-spin system, so for the pure exchange interaction, the spins at
two sublattices must obey the scalar form of spins, $\partial
S_1/\partial t =- \partial S_2/\partial t$, with the extra term from
demagnetization.  In our system, each spin is coupled with more than
four neighboring spins, so we take two neighboring spins as an
example. For the above nonswitchable case (Fig. \ref{fig2}), we find
that $S_{1x}+S_{2x}$ and $S_{1y}+S_{2y}$ are not constant, so they do
not obey $\partial S_1/\partial t = - \partial S_2/\partial t$.  Our
$S_{1x}+S_{2x}$ decreases from 0 to about $-1\hbar$ with oscillations,
while $S_{1y}+S_{2y}$ increases from 0 to about $+1\hbar$ at the same
rate. For our switchable case (Fig. \ref{fig1}), $\partial
S_1/\partial t =- \partial S_2/\partial t$ is not fulfilled
either. Instead, we find that our result obeys the vector form ${\bf
  S}_1(t)/|{\bf S}_1(0)|=-{\bf S}_2(t)/|{\bf S}_2(0)|$.  This shows
that the simple argument based on a two-spin model is not applicable
to our realistic case.  We plan to investigate this issue further in a
much larger system.

In conclusion, we have carried out a joint first-principles density
functional theory and model simulation of all-optical spin reversal in
Mn$_2$RuGa. We are able to find a case that the spins can be switched
without employing a magnetic field. The system also shows an optimal
laser electric field amplitude, with the same profile like those in
other systems.  The spins at Mn$_1$ site are switched from the $+z$
axis to $-z$ axis, while those at Mn$_2$ site are switched from the
$-z$ axis to $+z$ axis. This is fully consistent with the experimental
findings \cite{banerjee2019}. We find that the exchange interaction
has a significant effect on the switching. When we reduce the exchange
to 0.001 eV, we find the system becomes unswitchable. It behaves like
a regular antiferromagnet. Our finding is expected to motivate further
theoretical and experimental investigations in the future.

\acknowledgments This work was solely supported by the U.S. Department
of Energy under Contract No.  DE-FG02-06ER46304.  Part of the work was
done on Indiana State University's Quantum Cluster and High
Performance computers.  This research used resources of the National
Energy Research Scientific Computing Center, which is supported by the
Office of Science of the U.S.  Department of Energy under Contract
No. DE-AC02-05CH11231. Our calculations also used resources of the
Argonne Leadership Computing Facility at Argonne National Laboratory,
which is supported by the Office of Science of the U.S. Department of
Energy under Contract No.  DE-AC02-06CH11357.

{Availability of data.} The data that support the findings of this
study are available from the corresponding author upon reasonable
request.

https://orcid.org/0000-0002-1792-2701

\end{document}